# CLUSTERING OF DIRBE LIGHT AND IR BACKGROUND

A. Kashlinsky[1], J.C. Mather[2], S. Odenwald[2,3], M.G. Hauser[4]
[1]NORDITA, Blegdamsvej 17, Copenhagen, DK-2100 Denmark
[2]Code 685, Goddard Space Flight Center, Greenbelt, MD 20771 USA
[3]Applied Research Corporation
[4]Code 680, Goddard Space Flight Center, Greenbelt, MD 20771 USA

## ABSTRACT

We outline a new method for estimating the cosmic infrared background using the spatial and spectral correlation properties of infrared maps. The cosmic infrared background from galaxies should have a minimum fluctuation of the order of 10% on angular scales of the order of 1°. We show that a linear combination of maps at different wavelengths can greatly reduce the fluctuations produced by foreground stars, while not eliminating the fluctuations of the background from high redshift galaxies. The method is potentially very powerful, especially at wavelengths where the foreground is bright but smooth.

## 1. INTRODUCTION

Diffuse background radiation fields contain information about the entire history of the universe, including those periods in which no discrete objects exist or can be detected by telescopic study. The cosmic infrared background (CIB) should contain much of the luminosity of the early generations of stars and galaxies, since the cosmic expansion redshifts the wavelength of peak luminosity from these sources from the UV and visible bands into the infrared. In addition, dust absorbs much of the original UV luminosity and re-emits it in the IR. The CIB therefore contains important information about the evolution of the Universe between the redshift of last scattering (probed by the microwave background) and today (probed by optical surveys).

The CIB has still not been discerned in the sky brightness data measured by the COBE Diffuse Infrared Background Experiment (DIRBE), which was designed for the purpose, because it is hard to distinguish the CIB from foreground sources such as interplanetary and interstellar dust, and stars. Extrapolations of galaxy counts clearly predict a near IR CIB well above the DIRBE noise level. The most direct analysis of the DIRBE data gives the limits on the CIB derived from the darkest parts of the sky (Hauser 1995). At the DIRBE $J$, $K$, and $L$ bands they are respectively $\nu I_\nu = 393 \pm 13$, $150 \pm 5$, and $63 \pm 3$ nW m$^{-2}$sr$^{-1}$. The latest attempt by the DIRBE team to account for foreground emission is reported by Hauser (1995), giving residuals for these bands ranging from 50 to 104, 15 to 26, and 15 to 24 nW m$^{-2}$sr$^{-1}$ respectively.

An alternative detection method was suggested by Gunn (1965) based on the correlation properties of the diffuse backgrounds. It was successfully applied by Shectman (1973, 1974) in the $V$ band and Martin and Bowyer (1989) in the UV. Indeed, if it comes from the redshifted light emitted by galaxies, the diffuse background must show fluctuations to reflect the correlation properties of galaxies.

We present here an overview of this study: section 2 gives a brief theo-retical background for the method, data analysis is presented in section 3 and we conclude in section 4. More details and final numbers will be found in our forthcoming paper (Kashlinsky, Mather, Odenwald, and Hauser 1995; hereafter Paper I).

## 2. THEORETICAL PRELIMINARIES

If the primordial density fluctuations were Gaussian as most theories assume, then the three-dimensional spatial correlation function, $\xi(r)$ determines all properties of the primordial density field. $\xi(r)$ has now been measured to large distances ( $< 100 h^{-1}$Mpc ) in the $B_J$ (Maddox et al. 1990; hereafter the 'APM' survey) and $r$ bands (Picard 1991; hereafter the 'Palomar' survey). These estimates are consistent with other measurements of the mass distribution (cf. Kashlinsky 1994 and references therein), although there is no proof that all types of dark matter follow luminous matter. While deeper than before, the APM and Palomar surveys map the galaxy distribution at rather low redshifts (the APM survey limit is $b_J = 20.5$, and the Palomar survey limit is $r = 19$).

The three-dimensional correlation function is not directly observed in two dimensional maps. We therefore consider the projected two dimensional correlation function as a directly observable quantity. For diffuse light, the projected 2-D correlation function between fluxes in two different wavebands $\lambda_1, \lambda_2$ is defined as $C_{\lambda_1,\lambda_2}(\theta) \equiv \langle \lambda_1 \delta I_{\lambda_1}(\mathbf{x}) \cdot \lambda_2 \delta I_{\lambda_2}(\mathbf{x}+\theta) \rangle$, where $\delta I = I - \langle I \rangle$. It is related to the underlying 2-point correlation function via the analog of the Limber equation, which in the limit of small angles $\theta < 1$ reads:

$$C_{\lambda_1,\lambda_2}(\theta) = \int_0^\infty dz [\partial \lambda_1 I_{\lambda_1}/\partial z][\partial \lambda_2 I_{\lambda_2}/\partial z] \int_{-\infty}^\infty d\Delta \xi(r_{12}; z), \qquad (1)$$

where $\Delta \equiv (z_2 - z_1)$, $r_{12}^2 = (\Delta c dt/dz)^2 + x^2(z)/(1+z)^2 \theta^2$, and $x(z) = c \int_0^z (1+z) dt$ is the comoving distance to $z$.

The angular scale of the DIRBE beam probes comoving linear scales of $< 20 - 30 h^{-1}$Mpc for the range of redshifts contributing to the bulk of the (near) IR extragalactic flux. On these scales, the correlation function implied by the APM survey can be approximated as a power law of slope $\gamma + 1 \simeq 1.7$. Thus as $\theta \to 0$ we can model $\xi(r; z)$ as $\xi(r; z) = (r/r_*)^{-1-\gamma} \Psi^2(z)$, where $r_* = 5.5 h^{-1}$Mpc and the function $\Psi(z)$ accounts for the evolution of the clustering pattern, and is defined to be unity for $z = 0$. This leads to a predicted zero-lag correlation signal (mean square deviation) seen by DIRBE with the top-hat beam of:

$$C(0) = 1.6 \times 10^{-19} D Q_\lambda^2 \int \frac{[E(z) S(\lambda; z) \Psi(z)]^2}{[(1+z)^6 \sqrt{1+\Omega z}]} [\frac{R_H(1+z)}{x(z)}]^{0.7} dz \qquad (2)$$

where $C(0)$ is in W$^2$m$^{-4}$sr$^{-2}$, $Q_\lambda = \lambda f_\lambda(\lambda; 0)/\Delta \lambda_B f_\lambda(\lambda_B; 0)$. Here $\lambda_B$ and $\Delta \lambda_B$ refer to the $B$ band filter wavelength and bandwidth used to define the galaxy luminosity functions. For galactic spectra and galaxy type mixes adopted from Yoshii and Takahara (1988) we get $Q_J = 8.9$, $Q_K = 3.8$, $Q_L = 1.7$ for the $J, K, L$ bands. $D$ is the beam dilution factor which for circular top-hat beam and $\gamma = 0.7$ is $D(\gamma) \simeq 1.4$. We evaluated the effective radius of a circular beam corresponding to the DIRBE instrument to have radius $\simeq 0.45°$, after accounting for the effects of the beam shape, the beam scanning during the observation time, the pixelization process, and the attitude errors. $E(z)$ and $S(\lambda, z)$ are the evolution parameters and are defined in Paper I. Thus we expect the zero-lag signal to be near $C_\vartheta(0) \sim 10^{-2}(\lambda I_\lambda)^2$.

Even if the data do not allow a detection, this method allows translation of an upper limit on $C(0)$ into an upper limit on the total flux in that band. To see how the measurements of the correlation signals constrain the levels of the background it is illustrative to rewrite the above equation as:

$$C(0) = 3.6 \times 10^{-3} \int (d\lambda I_\lambda/dt)^2 \Psi^2(t)(R_H/r)^\gamma H_0^{-1} dt \qquad (3)$$

where $t$ is the cosmic time and $r = x(z)/(1+z)$. In the Friedman-Robertson-Walker Universe the ratio of $[r(z)/R_H]^\gamma$ reaches a maximum of 0.45 at $z \sim 1.5$. The value at the maximum is almost independent of $\Omega$ being 0.43 for $\Omega = 0.1$ and 0.47 for $\Omega = 1$. This translates into $[R_H/r(z)]^\gamma \geq 2.2$. Thus any detection of the cosmological $C(0)$ or an upper limit on it would set an upper limit on the rms $z$-gradient of the total flux of

$$\left(\frac{d\lambda I_\lambda}{dz}\right)_{\text{rms}} \equiv \sqrt{\int (d\lambda I_\lambda/dz)^2 dz} \leq 11\sqrt{C(0)}. \qquad (4)$$

To relate this $z$-gradient to the total flux requires a model for the $z$ dependence of $d\lambda I_\lambda/dz$. Reasonable classes of models based on recent galaxy formation ($z$ of a few) give the ratio of $\lambda I_\lambda$ to the $z$-gradient of the order of unity, and any specific model can be evaluated easily. This method allows for a simple and efficient way of constraining the CIB from material clustered like galaxies, based on the measurements of the mean square deviation in the DIRBE maps.

In the conventional DIRBE analysis one derives upper limits from the DC component of the signal, using various methods for modeling and subtracting the smooth and discrete foreground sources. As eq. (4) shows the correlations method could be quite efficient when the foreground emission is very bright but homogeneous. This is typically the case in wavebands longward of the ones considered here, where zodiacal light dominates stellar emission. The two methods measure different quantities, one depends on foreground models and one on background models, and both provide valuable constraints on the processes of galaxy formation, evolution, and spatial clustering. The analysis at these longer wavelengths will be presented in a future paper. Nevertheless, applying the techniques in the near-IR already gives interesting results.

## 3. DATA REDUCTION

We analyzed data from the three shortest wavelength bands observed by DIRBE, which approximate the ground-based photometric bands $J, K, L$, at 1.25, 2.2, and 3.5 microns. We selected four fields of 128×128 pixels, or 38.5 × 38.5°, located outside the Galactic plane, $|b^{II}| > 20°$. Two fields are located in the Ecliptic plane and two near the north and south Ecliptic poles. We also studied the same five fields that are used in setting the strongest current limits on CIB (Hauser 1995): four fields are 32 by 32 pixels in size (10° by 10°) and are centered on the two ecliptic (hereafter referred to as SEP and NEP) and two galactic (SGP and NGP) poles. The fifth field is 5° by 5° (16 by 16 pixels) and is centered on the Lockman hole (hereafter LH), the region of the minimum HI column density at $(l, b) = (148°, +53°)$ (Lockman et al 1986). We analyzed two versions of these maps. Map A was constructed by interpolating the data from 4 consecutive weeks of observation to a constant solar elongation of 90°, thus reducing the effects of the variation of the zodiacal light with time. Map B was the grand average of all 42 weeks of observations, as delivered to the National Space Science Data Center (Leisawitz 1995), with the DIRBE model of the zodiacal light subtracted directly (Reach *et al.* 1995; Franz *et al.* 1995). The model uses the time dependence of the measured sky brightness as the unique signature of the zodiacal light.

A measurement of $C(0)$ must satisfy three criteria to be considered a firm detection. First, it must be isotropic to a high degree. Second, it must have a level significantly above the random noise levels from the instrument and data processing. The spatial power spectrum of the fluctuations must be consistent with an origin outside the instrument and data reduction process. Third, it must be clear that it can not be produced by foreground sources. It must not be spatially correlated with foreground sources like the Galactic or ecliptic plane, or the local supercluster. Even without passing all these tests, the measured limits on $C(0)$ and their implications for the clustered part of the CIB are still very interesting as an example of this promising method.

We now describe two steps that were applied to reduce the effects of the zodiacal and stellar foregrounds from the DIRBE maps: discrete source masking, and color subtraction.

Except near bright objects like the Large Magellanic Cloud or the Galactic Plane, the appearance of each map in the selected $J, K, L$ bands is characterized by a large number of bright point sources, mostly known stars, situated on a smooth distribution. For purposes of estimating the properties of the CIB we must eliminate the contribution from the discrete foreground as well as possible. Each field was processed by an iterative background modelling and source removal program developed by the DIRBE team. The smooth component was modeled as a 4th-order polynomial surface, which was fit to the entire field after application of a 5 × 5-pixel spatial filter. Pixels with fluxes greater than $N_{\rm cut}$ standard deviations above the fitted background model were blanked, as were the surrounding 8 pixels. In other words, each star was assumed to occupy a 3 × 3-pixel region 0.97° square, a factor of ∼ 2 larger in area than the nominal DIRBE beam size of 0.7° square.

The procedure was repeated for the remaining non-blanked pixels in the map until no further points were masked by the algorithm. The background model was recomputed for the non-blanked pixels in the field, and the blanking process performed again at the same clipping threshold factor, $N_{\rm cut}$. Since the clipping threshold is defined by a factor relative to the standard deviation of the residuals of the remaining pixels, it changes with each iteration and is slightly different for different values of $N_{\rm cut}$. This process was iterated until no new peaks were identified in the current map iteration. Because this process does not actually subtract a background from the data at each step in the iteration, the final product is a sky map containing all of the original low spatial frequency background information, but in which bright points sources have been deleted by blanking their pixels. We tested cutoff thresholds of $N_{\rm cut} = 3.5, 5,$ and 7; at lower clipping thresholds we have too few pixels remaining to allow for a meaningful analysis.

After the masks were created individually for all three bands, a combined mask was made so that only those pixels which were considered valid for all three bands would be used for further analysis. This step guarantees that the data are consistent for all the bands and for the color-subtracted analysis to be described below.

We examined the histograms of the fluctuations for each field and map and value of $N_{\rm cut}$. We found the following expected effects. The histograms are quite skew, with a high-intensity tail falling off roughly as $dN/d\delta I \propto \delta I^{-5/2}$, in accord with the prediction for stars filling infinite space uniformly. The low-intensity side of each histogram drops abruptly, since it is very improbable for no stars to be found in the large DIRBE beamwidth. If the beamwidth were only 1 arcminute, the result would be very different. Histograms for Map A, for which the zodiacal light model has not been subtracted, show double peaks for regions near the ecliptic plane. In Map B, all the fields give histograms with roughly comparable widths, as expected if the stellar foreground fluctuations are roughly the same for regions all at high galactic latitude. The measured width of the central part of the distribution is moderately sensitive to the clipping threshold used. We find that $N_{\rm cut} = 3.5$ is the lowest value where enough pixels still remain for proper analysis, because each deleted pixel also removes its 8 neighbors. The measured dispersion or $C(0)$ is still far too anisotropic to be interpreted as cosmological, but can be used to give upper limits. To obtain much lower values would require very careful analysis of the observational noise, the stellar foreground fluctuation contributions, and the detailed response properties of our nonlinear analysis algorithms.

The signal from the foreground radiation may also be disentangled from that of galaxies by considering the properties of the foreground stellar emission that are different from the CIB. The extragalactic objects have somewhat different colors than the Galactic stars, and they are redshifted as well. Furthermore, galaxies of different morphological types have different stellar populations and this too should help distinguish the colors of the foreground emission from our Galaxy (type Sc) from those of more representative galaxies (E/S0, Sa, etc.) which are expected to dominate the CIB at $J$, $K$ and $L$-bands.

On the other hand the near-IR DIRBE maps exhibit a clear color with a relatively small dispersion. We define the color between any two bands, 1 and 2, as $\alpha_{12} = I_1/I_2$. Within each field the dispersion of colors is small, between 5% and 10%. The colors confirm that most of the light of the Galaxy in these bands comes from K and M giants (Arendt *et al.* 1994).

This empirical property of the constancy of color found in the DIRBE near-IR maps can be used to reduce the foreground contribution to $C(0)$ by taking the scaled difference of two maps. On the other hand, as discussed above most CIB in the near-IR is expected to come from galaxies whose spectra are significantly redshifted and thus the CIB is not expected to be washed out by this procedure. These observations are the basis for the color subtraction method we present below.

There are two *independent* contributions to the measured fluctuations in the linear combinations of the DIRBE maps: the one that comes from the *total* foreground (stars, zodiacal light, and cirrus), which we denote as $C_{*,\Delta}(0)$, and the one that comes from the extragalactic background which we label $C_{g,\Delta}(0)$. We use the $\Delta$ notation to indicate that these variances refer to the differences of maps at two different wavelengths. The total $C_\Delta(0)$ is the sum of these two components. Let us assume that the Galactic and zodiacal foreground component has the mean color between any two wavelengths 1 and 2 of $S_{12} = \langle \alpha \rangle$ and the dispersion in the color is $s_{12}^2 = \langle (\delta\alpha)^2 \rangle$, where $\delta\alpha$ is the deviation of the pixel color from the mean of the pixels in the field. Assuming that the color fluctuations are uncorrelated with the intensities, the zero-lag signal due to the foreground in Band 1, $C_{*,1}(0)$, can be written as

$$C_{*,1}(0) \simeq (S_{12}^2 + s_{12}^2)C_{*,2}(0) + s_{12}^2 \langle I_{*,2} \rangle^2. \qquad (5)$$

Note that the approximations used in this equation are not required for the validity of the analysis of the maps, but are only to indicate that major improvements are achievable with color subtraction. Let us construct the quantity which depends on the parameters in two bands (we define $\delta_i \equiv I_i - \langle I_i \rangle$), $\Delta_{*,12} \equiv \delta_{*,1} - \beta \delta_{*,2}$, where the quantity $\beta$ is for now an arbitrary number which will be quantified later. The zero-lag signal in the "color-subtracted" quantity defined above would be given by

$$C_{*,\Delta}(0) \equiv \langle \Delta_{*,12}^2 \rangle = [(S_{12} - \beta)^2 + s_{12}^2]C_{*,2}(0) + s_{12}^2 \langle I_{*,2} \rangle^2. \qquad (6)$$

It is clear from the above that if we were to choose $\beta$ close to the mean color for the foreground, $\beta \simeq S_{12}$, then the largest term in Eq.(6) could be made nearly zero and the contribution of the Galactic foreground confusion noise to the zero-lag signal $C_{*,\Delta}(0)$ in the difference map would be minimized. Furthermore, if $s \ll S$, that contribution can be decreased over the one specified in Eq.(5) by a fairly substantial amount.

Measurement shows that indeed we have $s \ll S$. Thus the variance in the color-subtracted fluctuation $\Delta$ due to the confusion noise can be reduced by a substantial factor, with a minimum when $\beta = S_{12}$. The plots of $C_\Delta(0)$ for the data from all five fields as a function of $\beta$ for $J-K$, $J-L$ and $K-L$ subtraction, for map B and $N_{cut} = 3.5$ show that the minimum of $C_\Delta(0)$ is indeed quite deep. Therefore, in constructing the color-subtracted quantities we use the values of $\beta$ for each field that correspond to the minimum of $C_\Delta(0)$. Unlike the single band zero-lag signal, which would correspond to $\beta = 0$, the minimum of $C_\Delta(0)$ is distributed significantly less anisotropically. The major anisotropy in the colour subtracted signal comes from the SEP field. This field, however, contains the LMC galaxy and it is fair to exclude it from further discussion of the colour subtracted signal.

The color subtraction works well to suppress the fluctuations due to the Galactic foreground. On the other hand the predicted extragalactic background fluctuations are unlikely to be removed by the color subtraction method because the redshifts give the background a different color than the foreground. The extragalactic signal would change by a different amount, and may even increase because of the combination of redshift and morphological effects. This can be shown explicitly using eqs.(2)-(4) and is discussed in more detail in our Paper I. Since the Galaxy colors are typical of general stellar populations our method effectively reduces the contribution from the nearby galaxies, but because of the redshift effects on galaxies the overall zero-lag signal remains at the levels of the single band cases. In other words, the color subtracted signal probes galaxies in a window of redshifts centered on an early epoch. The distant galaxies do not have a single color, and can not be removed by the color subtraction.

## 4. CONCLUSIONS

We have tested a promising method for recognizing the contribution of high-redshift galaxies to the cosmic infrared background, using their spatial and spectral correlation properties. These are different from the foreground stars, which have a single dominant color with small dispersion. Bright foreground stars can also be eliminated by simple masking techniques. We prepared linear combinations of maps at different wavelengths and adjusted the coefficient to minimize the resulting variance. The residuals are much smaller than those of the single band maps, and are much more isotropic. Theoretically, the fluctuations of the CIB should have different colors from the foreground stars and should still be detectable in these linear combination maps.

Even the single band fluctuation limits can be interpreted as (model-dependent) limits on the CIB originating in material clustered like galaxies. In the $J, K, L$ bands, with reasonable galaxy evolution models, they are significantly lower than direct estimates of the upper limits on the CIB derived from the DIRBE maps and are comparable to the residuals in the direct sky brightness analysis (Hauser 1995).

There is still work to do in identifying the sources for all of the measured fluctuations. We have demonstrated that the instrument noise is quite small in the single band maps, but not negligible in the analysis of the color subtracted maps. The possible sources for such fluctuations could be the long-sought CIB, the residuals of the foreground star fluctuations, or effects of patchy dust absorption. A spatial power spectrum analysis is needed to show which of the above contributes to the detected signal.

## REFERENCES

Arendt, R. G., *et al.*, 1994, Ap.J., 425, L85
Franz, B. A. *et al.*, 1995, this Workshop
Gunn, J. 1965, Ph.D.Thesis, Caltech. (unpublished)
Hauser, M. G, 1995, this Workshop
Kashlinsky, A. 1994, in "Evolution of the Universe and its Observational Quest", ed. K. Sato, p.181, Universal Academy Press, Inc. - Tokyo
Kashlinsky, A., Mather, J., Odenwald, S., and Hauser, M. 1995, Ap.J., to be submitted
Leisawitz, D. T., 1995, this Workshop
Lockman, F. J. *et al.*, 1986, Ap.J., 302, 432
Maddox, S. *et al.*1990, MNRAS, 242, 43p
Martin, C. and Bowyer, S. 1989, Ap.J, 338, 677
Picard, A. 1991, Ap.J., 368, L7
Reach, W. T. *et al.*, 1995, this Workshop
Shectman, S. 1973, Ap.J., 179, 681
Shectman, S. 1974, Ap.J., 188, 233
Yoshii, Y. and Takahara, F. 1988, Ap.J., 326, 1